\documentclass[reprint,amsmath,amssymb,showpacs,aps,]{revtex4-1}
\usepackage{graphicx}
\usepackage{dcolumn}
\usepackage{bm}
\usepackage{soul}
\usepackage{ulem}
\usepackage{color}
\usepackage[T1]{fontenc}
\usepackage[super]{nth}
\usepackage{amsmath}
\usepackage[center]{subfigure}
\usepackage[utf8]{inputenc}
\usepackage{picture}
\usepackage[super]{nth}
\usepackage{float}

\begin{document}
\title{Finite-Range Coulomb Gas Models of Banded Random Matrices and Quantum Kicked Rotors}
\author{Akhilesh Pandey}
\email{apandey2006@gmail.com, ap0700@mail.jnu.ac.in}
\affiliation{School of Physical Sciences, Jawaharlal Nehru University, New Delhi-110067, India}
\author{Avanish Kumar}
\email{avanishkumar31@gmail.com}
\affiliation{School of Physical Sciences, Jawaharlal Nehru University, New Delhi-110067, India}
\author{Sanjay Puri}
\email{purijnu@gmail.com}
\affiliation{School of Physical Sciences, Jawaharlal Nehru University, New Delhi-110067, India}
\date{\today}

\begin{abstract}
Dyson demonstrated an equivalence between infinite-range Coulomb gas models and classical random matrix ensembles for study of eigenvalue statistics. We introduce finite-range Coulomb gas (FRCG) models via a Brownian matrix process, and study them analytically and by Monte-Carlo simulations. These models yield new universality classes, and provide a theoretical framework for study of banded random matrices (BRM) and quantum kicked rotors (QKR). We demonstrate that, for a BRM of bandwidth $b$ and a QKR of chaos parameter $\alpha$, the appropriate FRCG model has the effective range $d=b^2/N= \alpha^2/N$, for large $N$ matrix dimensionality. As $d$ increases, there is a transition from Poisson to classical random matrix statistics.
\end{abstract}
\maketitle
\textit{Introduction.}\textemdash Random matrices are being used extensively in many branches of physics as well as in other disciplines. Specifically they have found applications in quantum chaotic systems, such as complex nuclei, atoms, molecules, mesoscopic systems, disordered systems and model quantum systems of few degrees of freedom \cite{CEP,EPW1,Brody,BGS,BG,ML,CC,CWB,GMW,AM,FH,CH,HS}. In recent years newer applications have emerged in biology \cite{RA,AST}, economics \cite{EC1,EC2} and in communication engineering \cite{COM1, COM2,COM3}. In most of these applications, classical ensembles (circular and Gaussian ensembles) have provided a basis for understanding statistical properties of complex spectra. \par 
{Many studies have focused on the crossover from Poisson to classical random matrix results as a physical parameter is varied. In this context, there have been studies of diverse systems such as atomic spectra \cite{RP}, random matrix models \cite{AP2},  quantum chaotic systems \cite{WM,SVZ,BR,FKPT,GH,AP1}, Anderson localization \cite{SM,BKS,MMMS}, quark-gluon plasma \cite{GKP} and neural networks \cite{AHN}. The crossover has been studied  extensively in model systems such as quantum kicked rotors (QKR) and banded random matrices (BRM) \cite{FM1,PF,FM2,CMI,CGIS,WFL,CIM,FYMR,FYMRL,CCGI,CGIlZ,FCIC}. There are several routes where by this transition can occur. In this letter, we provide a theoretical framework to obtain spectral properties for such a crossover in QKR and BRM.\par
Dyson has shown that joint probability distributions (jpd) of the classical ensembles are equilibrium states of Brownian motion model of $N$ Coulombic particles interacting with each other. The positions of the Brownian particles are identified as the eigenvalues of the random matrix problem \cite{FDY}. In his original papers Dyson considered Coulombic particles on a real line with harmonic binding (referred to as the Gaussian ensembles (GE)) and Coulombic particles on the unit circle (referred to as the \textit{circular ensembles} (CE)). In an important generalization Dyson introduced non-harmonic confining potentials on the real line \cite{FDY1}; see also \cite{AP1,GPPS,KP1,PPK}. We will refer to these as \textit{linear ensembles}. Similar generalizations have been done for circular ensembles \cite{KP2}.\par
In the Dyson model and the extensions above, all particles have pairwise interaction. In this Letter we consider natural generalizations to the case where particles have finite-range interactions. We present analytic and Monte-Carlo results for such ensembles. We refer to these ensembles as finite-range Coulomb gas (FRCG) models. Some of the FRCG analytic results reported here were obtained earlier by one of the authors \cite{AP} and have been used in \cite{BGS1,JK,BGS2}. The main purpose of this letter is to demonstrate that, for eigenvalue spectra, FRCG models are in one-to-one correspondence with BRM and QKR. We will show numerically that the range $d$ of the FRCG model equals $b^{2}/N=\alpha^{2}/N$. Here, $b$ is the bandwidth of the BRM, $\alpha$ is the `chaos parameter' of the QKR, and $N$ is the matrix dimensionality. We emphasize that the equality connecting BRM and QKR was discovered empirically by Casati, Izrailev and others \cite{FM2,CMI,CGIS,CIM}. Following Izrailev \cite{FM2}, we expect that the Brownian motion model described below for FRCG will have the same correspondence for eigenvectors. \par
\textit{FRCG models.}\textemdash We introduce the jpd of the eigenangles for the FRCG models of circular ensembles:
\begin{equation}\label{eq:1}
 P(\theta_1,\cdots,\theta_N) = C \exp(-\beta W).
\end{equation}
Here the $ \theta_{j}$ are the eigenangles (in ascending order), $C$ is the normalization constant, and $\beta$ is the Dyson parameter with values $1,2$ and $4$. The potential $W$ is given by 
\begin{equation}\label{eq:2}
W= -\sum {}^{'}\log|e^{i\theta_{j}}-e^{i\theta_{k}}|+ \sum V(e^{i\theta_{j}}),
\end{equation}
where $\sum{}^{'}$ denotes the sum over all $|j-k|\leq d$ with $j< k$. The logarithmic terms represent the finite-range repulsive Coulomb gas potential with range $d$, and $V$ is a one-body (real) periodic potential on the unit circle. Classical ensembles arise for $d=N-1$ and $V=0$, \textit{viz.,} circular orthogonal ensemble (COE), circular unitary ensemble (CUE) and circular symplectic ensemble (CSE) characterized by $\beta=1,2,4$ respectively. For the linear ensembles, we consider eigenvalues $ x_{j}$ instead of the $e^{i\theta_{j}}$ in (\ref{eq:1}, \ref{eq:2}) with $V$ a binding potential on real line. In this case for $d=N-1$ and $V$ harmonic, we recover the Gaussian orthogonal ensemble (GOE), Gaussian unitary ensemble (GUE) and Gaussian symplectic ensemble (GSE) for $\beta=1,2,4$ respectively.\par
\textit{Brownian banded matrix process for FRCG.}\textemdash The above jpd can be interpreted as the equilibrium density of a Brownian process with potential given in (\ref{eq:2}). This is a natural generalization of the Dyson Brownian motion model \cite{FDY,FDY1}. We demonstrate this for circular ensembles with $\beta=2$ and $V=0$. The corresponding Brownian matrix process involves a stochastic matrix increment via a BRM of bandwidth $d$. We introduce a Brownian process of unitary matrices $U(\tau)$:
\begin{equation}\label{eq:3}
U(\tau+\delta\tau)= U(\tau)\exp{\big(i\sqrt{\delta\tau} M(\tau)\big)}.
\end{equation}
Here, $\tau$ is a fictitious time, $\delta\tau$ is infinitesimal time and the Brownian increment is given in terms of hermitian random matrices $M(\tau)$, independent for each $\tau$. \textit{$M(\tau)$ is a banded matrix in $U(\tau)$-diagonal representation similar to the GUE matrix with the constraint that the nonzero matrix elements of $M_{jk}$ are only for $|j-k|\leq d$. Note that eigenangles $(\theta_{1},\cdots,\theta_{N})$ of $U(\tau)$ at each Brownian step are in ascending order. Thus $U(\tau)$ is related to $U(0)$ by a product of infinitesimal random perturbations. For eigenvector studies one should deal directly with the matrix ensembles $\lbrace U(\tau) \rbrace$.} \par
Using second-order perturbation theory \cite{FDY,SP1} we obtain for the eigenangles $\theta_{j}(\tau)$:
\begin{align}\label{eq:4}
\overline{\delta\theta_{j}} & = E(\theta_{j})\delta\tau, & \overline{(\delta \theta_{j})^{2}} & = 2\beta^{-1}\delta\tau.
\end{align}
Here the bar denotes the ensemble average and $E(\theta_{j})$ is given by $E(\theta_{j})=-\partial W/ \partial \theta_{j}$. This defines a Smoluchowski process with equilibrium density given in (\ref{eq:1}). The cases with $\beta=1,4$ can be dealt with similarly respecting the appropriate symmetries. Later we will make a correspondence between this Brownian motion process and a BRM ensemble with band-width $b\simeq\sqrt{Nd}$. Moreover we will also demonstrate a direct realization of (\ref{eq:3}) in QKR with the momentum operator playing the role of $M$. \par
\textit{Universality.}\textemdash We focus on $d = O(1)$, where we expect new universality classes of fluctuation properties for each $d$. To prove this we have used a moment method to obtain the eigenangle density (for large $N$) as:
\begin{equation}\label{eq:5}
\bar{\rho}{(\theta)}  \propto  \exp \left[-\beta V/(\beta d+1)\right].
\end{equation}
Using this we define the unfolded spacings $s_{j}= N|\theta_{j+1}-\theta_{j}| \bar{\rho}(\theta_{j})$. Thus the jpd, $ P_{d}(s_{1},\cdots,s_{N})$, of $N$ consecutive spacings is given in the circular case
\begin{equation}\label{eq:6}
P_{d} \propto \delta(\sum_{i=1}^{N}s_{i}-N) \prod_{j=1}^{N}\prod_{k=0}^{d-1}(s_{j}+\cdots+s_{j+k})^{\beta}.
\end{equation}
In the linear case, we again obtain (\ref{eq:6}), with the constraint $\sum s_{j}\lesssim N$, instead of the $\delta$-function term. For large $N$, the difference between the two cases can be ignored. Thus the same fluctuations are obtained for linear and circular ensembles, and the result is independent of the potential $V$. Each $d$ and $\beta$ gives rise to a unique class of fluctuations. As we will see later, these are realized in QKR and BRM. As $d$ increases, there will be a crossover from Poisson to classical (Wigner-Dyson) statistics. \par
We will derive analytic results from (\ref{eq:6}) and supplement them with Monte-Carlo (MC) calculations of (\ref{eq:1}). Our MC approach follows that of \cite{GPPS} for the linear case, and \cite{KP2} for the circular case. Note that, because of the logarithmic singularity in (\ref{eq:2}), the eigenangles (and the eigenvalues in the linear case) do not change their order.\par
\textit{Fluctuation measures.}\textemdash The statistical quantities of interest are as follows \cite{Brody,BG,ML}. For $n\ll N$, the jpd of $n$ consecutive spacings $P^{(n)}_{d}(s_{1},\cdots,s_{n})$ is obtained from Eq.~(\ref{eq:6}) by integrating over other variables $(s_{n+1},\cdots,s_{N})$ for $N \rightarrow \infty$. 
\begin{equation}\label{eq:7}
\begin{split}
P^{(n)}_{d}
\equiv\lim_{N\to\infty}  \int_{0}^{\infty}\cdots \int_{0}^{\infty}P_{d}(s_{1},\cdots,s_{N})ds_{n+1}\cdots ds_{N}.
\end{split}
\end{equation}
The \textit{spacing density} is given by $(n\geq 1)$
\begin{equation}\label{eq:8}
\begin{split}
p_{n-1}(s)=\int_{0}^{\infty}\cdots\int_{0}^{\infty} \delta (s-\sum_{j=1}^{n}s_{j})
P_{d}^{(n)}ds_{1}\cdots ds_{n}.
\end{split}
\end{equation}
The \textit{spacing variance} is denoted as $\sigma^{2}(n-1)$. The two-level correlation and cluster functions $R_{2}(s)$ and $Y_{2}(s)$ are:
\begin{equation}\label{eq:9}
R_{2}(s)= 1- Y_{2}(s)= \sum_{n=1}^{\infty}p_{n-1}(s).
\end{equation}
The \textit{number variance} $\Sigma{}^{2}(r)$ is the variance of the number of levels in intervals of fixed length $r$, and is given by,
\begin{equation}\label{eq:10}
\Sigma{}^{2}(r)= r-2\int_{0}^{r}(r-s)Y_{2}(s)ds.
\end{equation}
We show below that $Y_{2}(s)$ decays exponentially (or faster) for our FRCG models for $d=O(1)$. In that case $\Sigma^{2}(r)$ grows linearly with $r$, as shown in Eq.~(\ref{eq:19}). This should be contrasted with the logarithmic growth of $\Sigma^{2}(r)$ found in Gaussian ensembles.\par
\textit{Exact results for d=0,1.}\textemdash The cases $d=0,1$ are simple to deal with. The $d=0$ case corresponds to the Poisson ensemble. The corresponding jpd is the $d=0$ version of Eq.~(\ref{eq:6}), $P_{0}= C~\delta\Big( \sum_{i=1}^{N}s_{i}-N \Big)$. After $N-n$ integrations we obtain
\begin{equation}\label{eq:11}
P_{0}^{(n)}(s_{1},\cdot,s_{n})= \tilde{C} \Big(N-\sum_{i=1}^{n}s_{i}\Big)^{N-n-1} \overset{N\rightarrow\infty}{=}\bar{C} \exp(-\sum_{i=1}^{n}s_{i}).
\end{equation}
Thus using Eqs.~(\ref{eq:8}-\ref{eq:10}) we find 
\begin{align}\label{eq:12}
p_{n-1}(s)&= \frac{s^{n-1}}{(n-1)!}e^{-s}, & \sigma^{2}(n-1)&=n, & \Sigma^{2}(r)=r.
\end{align}
The cluster function $Y_{2}(s)=0$ for all $\beta$.
\par
For $d=1$, a similar calculation gives
\begin{eqnarray}\label{eq:13}
P_{1}^{(n)}(s_{1},\cdots,s_{n})= \tilde{C} \Big(N-\sum_{i=1}^{n}s_{i}\Big)^{N-n-1}\prod_{j=1}^{n}s_{j}^{\beta}\nonumber \\
\overset{N\rightarrow\infty}{=}\bar{C}\prod_{j=1}^{n}s_{j}^{\beta}e^{-(\beta+1)s_{j}},~~ \bar{C} =\frac{(\beta+1)^{n(\beta+1)}}{(\beta+1)!^{n}}.
\end{eqnarray}
Thus
\begin{eqnarray}\label{eq:14}
p_{n-1}(s)&=& \frac{(\beta+1)^{n(\beta+1)}}{n!(\beta+1)!}s^{n(\beta+1)-1}e^{-(\beta+1)s},\nonumber \\
  \sigma^{2}(n-1)&=&\frac{n}{\beta+1},~\Sigma^{2}(r)=\frac{r}{\beta+1}+\frac{\beta(\beta+2)}{6(\beta+1)^{2}}.
\end{eqnarray}
The cluster function $Y_{2}$ falls off exponentially, for example, $Y_{2}(s)= e^{-4s}$ for $\beta=1$.
\par
\textit{Mean-Field approximation.}\textemdash For other values of $d \ll N$  a mean-field (MF) calculation, in analogy with statistical physics, gives a good approximation to the above quantities. In Eq.~(\ref{eq:6}) we set $s_{j+1},s_{j+2},\cdots, s_{j+d-1}\simeq s_{j}$ in each of the $j^{th}$ factors, neglecting fluctuations in neighbouring spacings. Then integrating over variables $s_{n+1},\cdots,s_{N}$ this approximation yields
\begin{eqnarray}\label{eq:15}
P_{d}^{(n)}(s_{1},\cdots,s_{n})= \tilde{C} \Big(N-\sum_{i=1}^{n}s_{i}\Big)^{N-n-1}\prod_{j=1}^{n}s_{j}^{\beta}\nonumber \\
\overset{N\rightarrow\infty}{=}\left[\xi^{\xi}(\xi!)^{-1}\right]^{n} \prod_{j=1}^{n} s_{j}^{\beta}e^{-\xi s_{j}}, ~~ \xi= (\beta d+1),
\end{eqnarray}
\textit{i.e.}, the spacings are finally independent. This approximation is valid for $n \gtrsim d$. Eq. (\ref{eq:15}) is exact for $d=0,1$. Substituting this in (\ref{eq:8}), we obtain the spacing density for $n\geq1$
\begin{equation}\label{eq:16}
p_{n-1}(s)= \left[(n\xi-1)!\right]^{-1} \xi^{n\xi} s^{n\xi -1}e^{-\xi s}.
\end{equation}
In Fig.~\ref{fig:F1} we plot $p_{5}(s)$ for $d=2$ and $\beta=1, 2$. There is an excellent agreement between the MF approximation and MC numerical results.\par
The spacing variance is given by
\begin{equation}\label{eq:17}
\sigma^{2}(n-1)={n}{\xi}^{-1} + \gamma(\beta,d).
\end{equation}
The leading term in (\ref{eq:17}) (i.e., the linear term in $n$) is determined by the MF result (\ref{eq:16}). The constant term $\gamma(\beta,d)$ does not arise in the MF approximation, and is motivated by our exact $d=2$ result below. For $d>2$, we estimate this constant from our MC calculations. To calculate the number variance, we use (\ref{eq:15}, \ref{eq:16}) to obtain the Laplace transform of the two-level correlation function $R_{2}(s)$ in (\ref{eq:9}). We find for $\alpha\geq 0$,
\begin{equation}\label{eq:18}
\frac{1}{\alpha}-\int_{0}^{\infty}e^{-\alpha s}Y_{2}(s)ds = {\xi}^{\xi}/[({\xi+\alpha})^{\xi}-{\xi}^{\xi}].
\end{equation}
\begin{figure}[h!]
\begin{center}
\includegraphics[width=\linewidth, width=8.0cm]{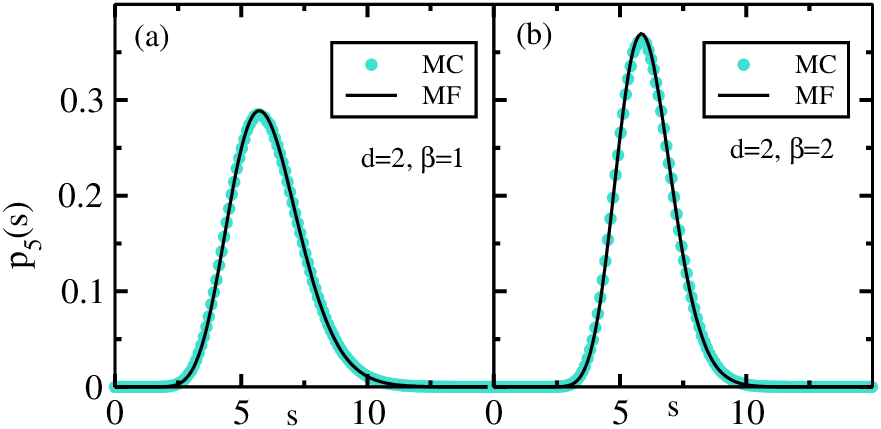}
\caption{Spacing density $p_{5}(s)$ for $d=2$ for (a) $\beta=1$ and (b) $\beta=2$.}\label{fig:F1}
\end{center}
\end{figure}
The cluster function $Y_{2}$ falls off exponentially for all $d$. Using (\ref{eq:18}) we can evaluate the number variance $\Sigma^{2}(r)$ in (\ref{eq:10}) as
\begin{equation} \label{eq:19}
\Sigma^{2}(r)= \sigma^{2}(r-1) + (\xi^2 -1)/(6\xi^{2}).
\end{equation}\par
\textit{Exact results for $d=2$.}\textemdash  We now outline a general framework to obtain exact results for $d>1$. For simplicity we focus on $d=2$. Our solutions are expressed in terms of the eigenvalues $\lambda_{\mu}$ and eigenfunctions $f_{\mu}$ of the integral equation 
\begin{equation}\label{eq:20}
\int_{0}^{\infty}e^{-(2\beta+1)s} s^{\beta}(s+t)^{\beta}f_{\mu}(s)ds=\lambda_{\mu}f_{\mu}(t).
\end{equation}
Here $\mu= 0,1,\cdots, \beta$. (A hermitian form of this equation is used in \cite{BGS2}). We order the eigenvalues as $\lambda_{0}<\lambda_{1}< \cdots < \lambda_{\beta}$. The $f_{\mu}$ are polynomials of order $\beta$. The largest eigenvalue $\lambda_{\beta}$ is positive and the corresponding eigenfunction has all positive coefficients. The $f_{\mu}$ satisfy the orthonormality relation
\begin{equation}\label{eq:21}
\int_{0}^{\infty}e^{-(2\beta+1)s} s^{\beta}f_{\mu}(s)f_{\nu}(s)ds= \delta_{\mu\nu}.
\end{equation}
By integrating Eq.~(\ref{eq:6}) over intermediate spacings from $s_{n+1},\cdots,s_{N}$ we find
\begin{align}\label{eq:22}
P_{2}^{(n)}&\propto \exp\left(-(2\beta+1)\sum_{i=1}^{n}s_{i}\right)\prod_{j=1}^{n}s_{j}^{\beta}\prod_{k=1}^{n-1}(s_{k}+s_{k+1})^{\beta}G(s_{1},s_{n}),
\end{align}
where $G$ is only a function of $s_{1},s_{n}$. It can be shown that $G$ is factorizable as $F(s_{1})F(s_{n})$ with $F(s) \propto f_{\beta}$. Thus $P_{2}^{(n)}$ can be written as
\begin{eqnarray}
\label{eq:23}
&& P_{2}^{(n)}(s_{1},\cdots s_{n})= \frac{1}{\lambda_{\beta}^{n-1}}\exp\left(-(2\beta+1)\sum_{j=1}^{n}s_{j}\right) \times 
\nonumber \\
&& ~~~~~~~~~~~~~~~ \prod_{j=1}^{n}s_{j}^{\beta}\prod_{k=1}^{n-1}(s_{k}+s_{k+1})^{\beta}f_{\beta}(s_{1})f_{\beta}(s_{n}),
\end{eqnarray}
valid for $n=1,2,3,\cdots$. Using Eq.~(\ref{eq:8}) one can obtain $p_{n-1}$. In particular the nearest-neighbor spacing density is $p_{0}(s)= e^{-(2\beta+1)s}s^{\beta}(f_{\beta}(s))^{2}$. \par
Integrating Eq.~(\ref{eq:23}) over all variables except $s_{1}$ and $s_{n}$ we find the jpd of $s_{1}$ and $s_{n}$ for $n>1$,
\begin{eqnarray}\label{eq:24}
Q_{n}(s_{1}, s_{n})= e^{-(2\beta+1)(s_{1}+s_{n})}s_{1}^{\beta}s_{n}^{\beta}f_{\beta}(s_{1})f_{\beta}(s_{n})\nonumber \\
\sum_{\mu=0}^{\beta}(\frac{\lambda_{\mu}}{\lambda_{\beta}})^{n-1}f_{\mu}(s_{1})f_{\mu}(s_{n}).
\end{eqnarray}
Using Eq.~(\ref{eq:24}) the covariance between $s_{1}$ and $s_{n}$ is given by
\begin{equation}\label{eq:25}
\langle s_{1}s_{n}\rangle -1 = \sum_{\mu=0}^{\beta -1}\left(\frac{\lambda_{\mu}}{\lambda_{\beta}}\right)^{n-1}\langle sf_{\beta}f_{\mu}\rangle^{2},
\end{equation}
and hence the spacing variance can be shown to be
\begin{eqnarray}
\label{eq:26}
&& \sigma^{2}(n-1) = n\left[\langle s^{2}f_{\beta}^{2}\rangle -1 + 2\sum_{\mu=0}^{\beta -1} \left(\frac{\lambda_{\mu}}{\lambda_{\beta}-\lambda_{\mu}} \right) \langle sf_{\beta}f_{\mu}\rangle ^2 \right] \nonumber \\ 
&& ~~~~~~~~ -2\sum_{\mu=0}^{\beta-1} \frac{\lambda_{\mu}\lambda_{\beta}}{(\lambda_{\beta}-\lambda_{\mu})^{2}}
\langle sf_{\beta}f_{\mu}\rangle ^2 \left[1-\left(\frac{\lambda_{\mu}}{\lambda_{\beta}} \right)^{n}\right].
\end{eqnarray}
Here the angular brackets represent
\begin{equation}\label{eq:27}
\langle F(s) \rangle = \int_{0}^{\infty}e^{-(2\beta+1)s}s^{\beta}F(s)ds.
\end{equation}
The term $\left(\lambda_{\mu}/\lambda_{\beta}\right)^{n}$ on the right hand side of (\ref{eq:26}) decays exponentially yielding (\ref{eq:17}) with $\xi=2\beta +1$.\par 
For $\beta=1$ we can obtain the analytic solution of the integral equation (\ref{eq:20}). The two eigenvalues and eigenfunctions are given 
respectively by $(\mu=0,1)$,
\begin{align}\label{eq:28}
\lambda_{\mu}&=\left(1\pm\sqrt\frac{3}{2}\right)\frac{2}{27}, & f_{\mu}(s)&= \frac{\left(s\pm\sqrt{\frac{2}{3}} \right)}{\sqrt{2|\lambda_{\mu}|\sqrt{\frac{2}{3}}}},
\end{align}
where $+$ and $-$ correspond to $\mu=1, 0$ respectively.
Using (\ref{eq:25}, \ref{eq:27}) we obtain (\ref{eq:18}) with $\gamma(1,2)=1/18$.
For $\beta=2,4$, numerical integrations yield $\gamma= 0.045157, 0.030598$  respectively.\par
\textit{Cases with $d>2$.}\textemdash The integral equation (\ref{eq:20}) can be generalized for $d>2$, and has $\zeta+1$ eigenvalues and eigenfunctions, where $\zeta=\beta(d-1)$. However the calculations become less tractable for larger $d$, and we use MC calculations to obtain a complete picture for arbitrary $d$. These results are shown in Fig.~\ref{fig:F2}, \ref{fig:F3}. As $d$ increases, there is a cross-over to the classical ensemble results. For example, for $\Sigma^{2}(r)$, the classical ensemble results are obtained for $r\ll d$. \par
\textit{BRM and QKR.}\textemdash First we introduce QKR which have been studied extensively as model systems of quantum chaos. Following Izrailev \cite{FM1}, we consider an $N$-dimensional matrix model for QKR. The evolution operator is given by $U=BG$, where $B(\alpha)= \exp\left[-i\alpha \cos(\varphi +\varphi_{0})/ \hbar \right]$, and $G=\exp\left[-i(p+\gamma)^{2}/2\hbar \right]$. Here $\varphi$ and $p$ are the position and momentum operators. Further $\alpha$ is the kicking parameter, $\varphi_{0}$ is the parity-breaking parameter and $\gamma$ is the time-reversal-breaking parameter. In the position representation
\begin{eqnarray}
\label{eq:29}
U_{mn} &=& \frac{1}{N}\exp\left[-i{\alpha}\cos\left(\frac{2\pi m}{N}+\varphi_{0}\right)\right] \times \nonumber \\ 
&& \sum_{l=-N^{\prime}}^{N^{\prime}}\exp\left[-i\left(\frac{l^{2}}{2}-\gamma l-\frac{2 \pi \mu l}{N}\right)\right],
\end{eqnarray}
where $\mu= m-n; m,n=-N^{\prime},-N^{\prime}+1,\cdots,N^{\prime}; N^{\prime}=(N-1)/2$. For large $\alpha$ the eigenvalue spectra of $U$ accurately exhibit classical ensemble fluctuations (e.g., spacing distribution, number variance etc.). These are characteristic of the $\beta=1$ case for $\gamma=0$ and the $\beta=2$ case for $\gamma \neq 0$, both with $\varphi_{0}$ not equal to zero. For our numerical simulations we choose $\varphi_{0}=\pi/2N$ and $\gamma=0,0.7$, the latter respectively for $\beta=1,2$.\par
Next we consider BRM ensembles $\lbrace{A}\rbrace$ of bandwidth $b$ as mentioned above. The jpd of the matrix distribution is $P(A) \propto \exp(-\textrm{tr} A^2/4v^{2}) $ with $v^{2}$ being the variance of the non-zero off-diagonal matrix elements. As in Eq.~(\ref{eq:3}) we consider a banded matrix $A$ such that the only non-zero elements $A_{jk}$ arise for $|j-k|\leq b$. The matrices are real-symmetric, complex hermitian and quaternion real self-dual for $\beta=1,2,4$ respectively. These ensembles have semicircular density for $b\gg 1$ \cite{AP2,WFL,CCGI}:
\begin{align}\label{eq:30}
\rho(x)&=2\sqrt{R^{2}-x^{2}}/\pi R^{2}, &  R^{2}&=8\beta b v^{2}.
\end{align}
\par
\begin{figure}
\begin{center}
\includegraphics[width=\linewidth, width=8.0cm]{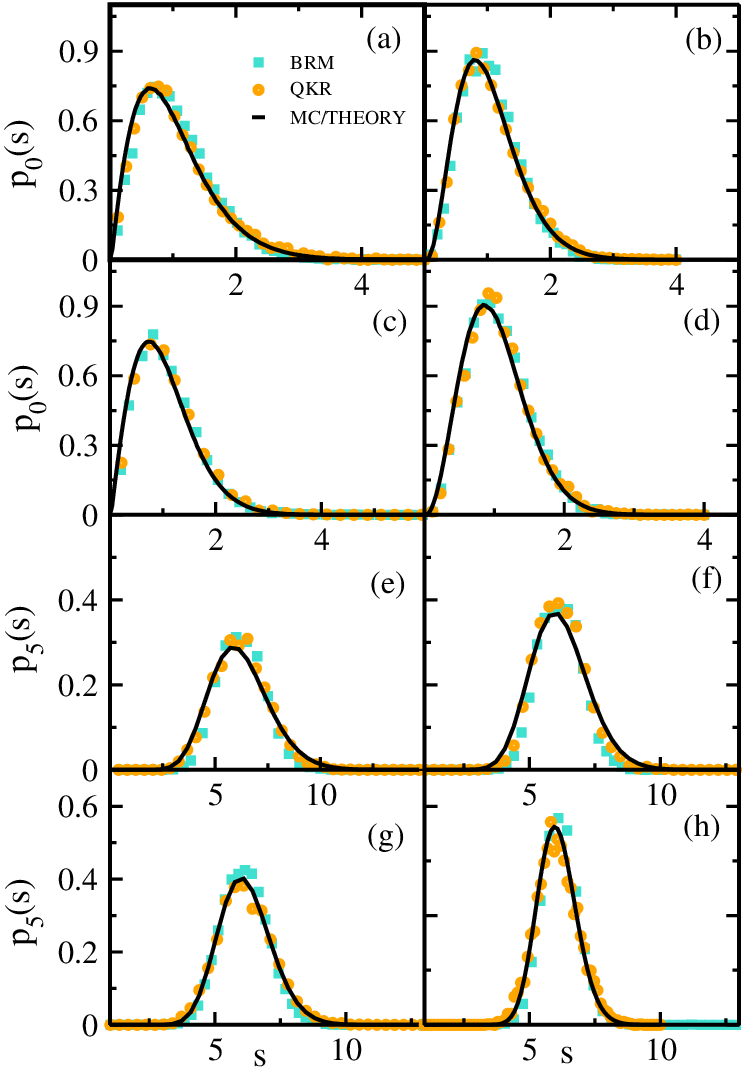}
\caption{Spacing density $p_k(s)$ for $k=0,5$: Left and right panels are for $\beta=1$ and $\beta=2$ respectively. (a), (b), (e), (f) correspond to $d=2$; (c), (d), (g), (h) correspond to $d=5$.}
\label{fig:F2}
\end{center}
\end{figure}
\begin{figure}
\begin{center}
\includegraphics[width=\linewidth, width=6.5cm]{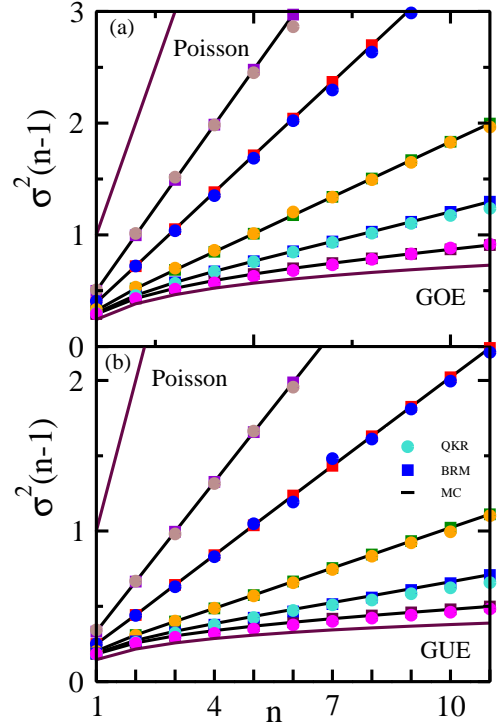}
\caption{Spacing variance for to  $d=1,2,5, 10, 25$ (from top to bottom) for (a) $\beta=1$ and (b) $\beta=2$. Poisson and GE results are also plotted. The MC results have been connected by a line for visual convenience.}
\label{fig:F3}
\end{center}
\end{figure}
\begin{figure}
\begin{center}
\includegraphics[width=\linewidth, width=7.25cm]{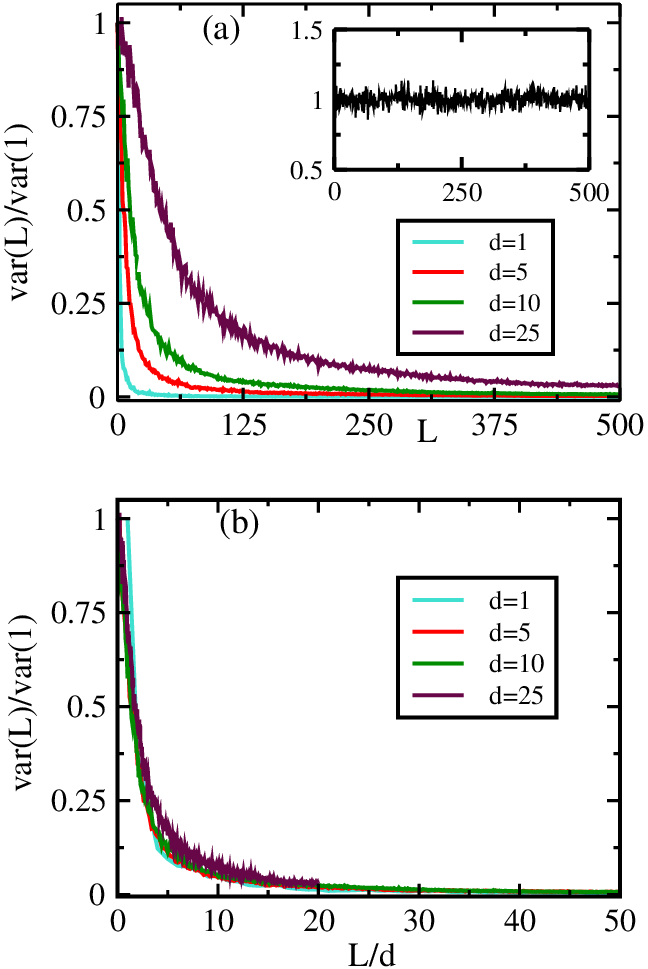}
\caption{Plot of var$(L)$/var$(1)$ vs. $L$ and scaled variable $L/d$, in (a) and (b) respectively, both for QKR. The inset in (a) is for large $\alpha$ such that $d\simeq O(N)$. We have taken $N=1000$.}\label{fig:F4}
\end{center}
\end{figure}
It has been shown \cite{CMI,CGIS,CIM,FM2}, that BRM and QKR give the same nearest-neighbor spacing density $p_{0}$(s) when $b^{2}/N$ and $\alpha^{2}/N$ are equal. One of the central results of this Letter is that BRM and QKR can be modeled by FRCG with $d= b^{2}/N=\alpha^{2}/N$ for all fluctuation measures. This is confirmed by our numerical results. As $d$ increases, there is a transition from the Poisson ensemble to the classical ensembles. \par
\textit{Generalizations to non-integer $d$.}\textemdash Our FRCG models above correspond to the case of integer $d$. To cover the entire parameter range we also introduce FRCG with non-integer $d$. In this case we generalize FRCG by modifying (\ref{eq:2}) so that (\ref{eq:6}) becomes
\begin{equation}\label{eq:31}
P_{d}=C\delta(\sum_{i=1}^{N}s_{i}-N) \prod_{j=1}^{N}\prod_{k=0}^{[d]}(s_{j}+\cdots+s_{j+k})^{\beta\Delta(k)},
\end{equation}
where $[d]$ is the largest integer $\leq d$. Moreover, $\Delta(k)=1$ for $k=0,1,\cdots, [d]-1$, and $\Delta([d])=d-[d]$. \par
\textit{Numerical results.}\textemdash We have performed extensive MC simulations of the FRCG, as well as simulations of BRM and QKR. We show representative numerical results here. For the MC calculations, $N=1000$ and data is averaged over $1000$ independent runs. Calculations for BRM and QKR are based on a single realization with $N=10000$ and $5000$ respectively. The values of $b$ and $\alpha$ are chosen to be $\sqrt{Nd}$. We have performed calculations for many values of $d$ including the non-integer cases (not shown here). In Fig~\ref{fig:F2}, we show results for $\beta=1, 2$ and $d=2, 5$ for $p_{k}(s)$ with $k=0, 5$. For $d=2$ the theory is derived from the integral equation in (18).  $p_{0}(s)$, $p_{5}(s)$ for QKR and BRM are in very good agreement with the FRCG results. In Fig~\ref{fig:F3}, we show $\sigma^{2}(n-1)$ vs. $n$  for $\beta=1,2$ and several $d$ values. Again the agreement between FRCG and BRM, QKR is excellent. We have also performed calulations for $\Sigma^{2}(r)$ and obtained similar agreement. The $d=50$ case (not shown here) is numerically coincident with GE results.\par 
\textit{Emergence of bandedness in QKR.}\textemdash Now we demonstrate that banded perturbations arise naturally in physical applications. In Fig.~\ref{fig:F4}(a), we show var$(L)$/var$(1)$ vs. $L$ for QKR. Var$(L)$ is the variance of the matrix elements, $p_{jk}$, of the momentum operator in the representation in which $B^{1/2}GB^{1/2}$ is diagonal \cite{PRS,SP}. $L=|j-k|$ is the distance from diagonal. var$(L)$/var$(1)$ decays rapidly with a scale depending on $d$. This rapid decay is a consequence of localization of eigenfunctions, analogous to the well studied phenomenon of Anderson localization \cite{PF,CMI,CGIS}. In Fig.~\ref{fig:F4}(b), we show the same quantity plotted against $L/d$. The excellent data collapse demonstrates that $d$ constitutes a natural scale for the decay of var$(L)$ as expected in banded matrices. For large $\alpha$, we obtain extended states, and the corresponding variance is independent of $L$ as shown in the inset of Fig.~\ref{fig:F4}(a). In our FRCG model we use a sharp cut-off with distance from diagonal rather than the exponential-like decay seen in Fig.~\ref{fig:F4}. This suggests that our results are not sensitive to the precise shape of the decay. \par
\textit{Conclusion.}\textemdash In summary, we have generalized Dyson's Brownian motion model to introduce finite-range Coulomb gas models. These models are solvable and we have presented detailed analytical results for spectral properties. The analytic results, supplemented by MC results, provide a comprehensive solution of the model. Further we have applied these models to understand spectral properties of BRM and QKR. There has been extensive study of both these systems. In this Letter we have provided a theoretical framework to compute their statistical properties. Finally, we note that in QKR all our results are applicable in the strongly chaotic regime, suggesting new universality classes of quantum chaotic systems.

\end{document}